\documentclass[useAMS,usenatbib]{mn2e}
\usepackage{graphicx}
\usepackage[dvips]{color}
\usepackage{txfonts}
\usepackage{amssymb}
\usepackage{natbib}

\title[Gravity modes in the massive binary V380\,Cyg]{Detection of
gravity modes in the massive binary V380\,Cyg
  from {\it Kepler} spacebased photometry and high-resolution
  spectroscopy\thanks{Based on DDT data gathered with NASA's Discovery mission
    {\it Kepler} and with the HERMES spectrograph, installed at the Mercator
    Telescope, operated on the island of La Palma by the Flemish Community, at
    the Spanish Observatorio del Roque de los Muchachos of the Instituto de
    Astrof\'{\i}sica de Canarias and supported by the Fund for Scientific
    Research of Flanders (FWO), Belgium, the Research Council of K.U.Leuven,
    Belgium, the Fonds National de la Recherche Scientific (F.R.S.--FNRS),
    Belgium, the Royal Observatory of Belgium, the Observatoire de Gen\`eve,
    Switzerland and the Th\"uringer Landessternwarte Tautenburg, Germany.}}

\author[A. Tkachenko et al.]{A.\ Tkachenko,$^1$  C.\ Aerts,$^{1,2,3}$ K.\ Pavlovski,$^4$
  J.\ Southworth,$^5$ P.\ Degroote,$^1$ J.\ Debosscher,$^1$ \newauthor M.\ Still,$^6$ S.\ Bryson,$^6$
  G.\ Molenberghs,$^3$ S.\ Bloemen,$^1$ B.~L.\ de Vries,$^1$ M.\ Hrudkova,$^{7,8}$ \newauthor R.\ Lombaert,$^1$
  P.\ Neyskens,$^9$ P.\ I.\ P\'{a}pics,$^3$ G.\ Raskin,$^1$ H.\ Van Winckel,$^1$ R.~L.\
  Morris,$^{10}$ \newauthor D.~T.\ Sanderfer,$^{11}$ and S.~E.\ Seader$^{10}$ \\
  $^1$Instituut voor Sterrenkunde, KU Leuven, Celestijnenlaan 200D, B-3001 Leuven, Belgium\\
  $^2$Department of Astrophysics, IMAPP, Radboud University Nijmegen, 6500 GL Nijmegen, The Netherlands\\
  $^3$Center for Statistics (CenStat), University of Hasselt, Agoralaan 1, B-3590 Diepenbeek, Belgium\\
  $^4$Department of Physics, University of Zagreb, Bijeni\v{c}ka cesta 32, 10000 Zagreb, Croatia\\
  $^5$Astrophysics Group, Keele University, Staffordshire ST5 5BG, UK\\
  $^6$NASA Ames Research Center/Bay Area Environmental Research Institute, MS 244-30, Moffett Field, CA 94 035, USA\\
  $^7$Th\"{u}ringer Landessternwarte Tautenburg, 07778 Tautenburg, Germany\\
  $^8$Isaac Newton Group of Telescopes, Apartado de Correos 321, E-387 00 Santa Cruz de la Palma, Canary Islands, Spain\\
  $^9$ Institut d'Astronomie et d'Astrophysique, Universit\'e Libre de Bruxelles, CP 226 Brussels, Belgium\\
  $^{10}$ SETI Institute/NASA Ames Research Center, Moffett Field, CA 94035\\
  $^{11}$NASA Ames Research Center, Moffett Field, CA 94035}

\date{Received date; accepted date}

\pagerange{\pageref{firstpage}--\pageref{lastpage}} \pubyear{2011}

\def\LaTeX{L\kern-.36em\raise.3ex\hbox{a}\kern-.15em
    T\kern-.1667em\lower.7ex\hbox{E}\kern-.125emX}

\begin{document}

\label{firstpage}

\maketitle

\begin{abstract}
  We report the discovery of low-amplitude gravity-mode oscillations in the
  massive binary star V380\,Cyg, from 180\,d of {\it Kepler} custom-aperture
  space photometry and 5 months of high-resolution high signal-to-noise
  spectroscopy. The new data are of unprecedented quality and allowed to improve
  the orbital and fundamental parameters for this binary.  The orbital solution
  was subtracted from the photometric data and led to the detection of periodic
  intrinsic variability with frequencies of which some are multiples of the
  orbital frequency and others are not.  Spectral disentangling allowed the
  detection of line-profile variability in the primary. With our discovery of
  intrinsic variability interpreted as gravity mode oscillations, V380\,Cyg
  becomes an important laboratory for future seismic tuning of the near-core
  physics in massive B-type stars.
\end{abstract}



\section{Introduction}

\begin{figure*}
\begin{center}
\rotatebox{270}{\resizebox{10cm}{!}{\includegraphics{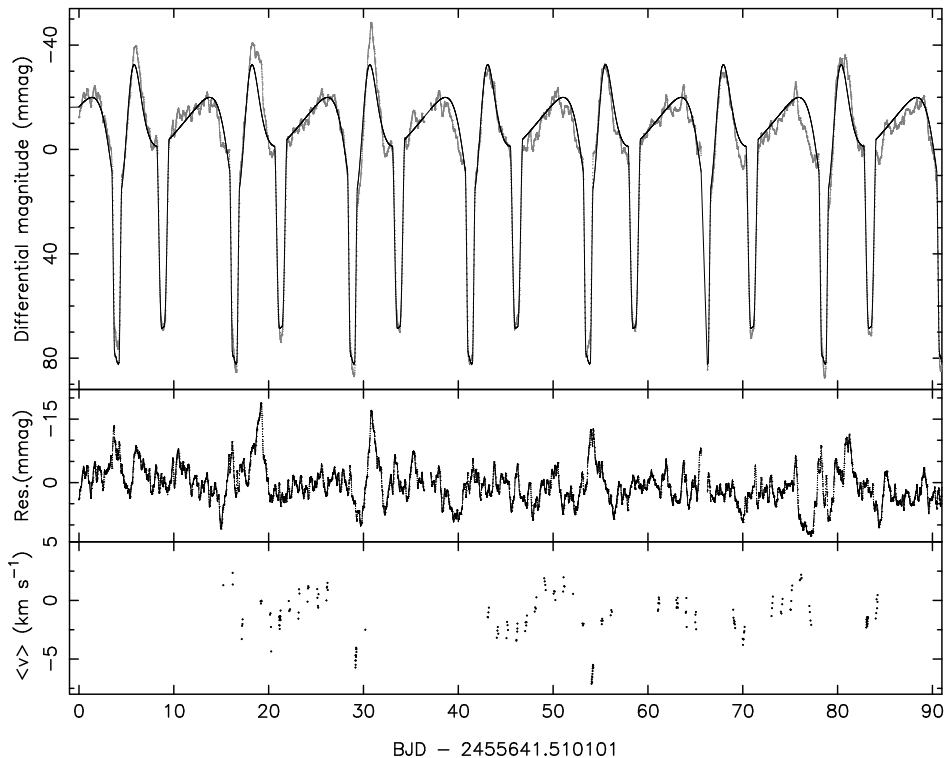}}}
\end{center}
\caption{Upper panel: {\it Kepler} light curve of V380\,Cyg observed during Q9
  (gray) with the binary model overplotted (black). Middle panel: residual light
  curve. Lower panel: first moment of the disentangled Si\,{\small III} line of
  the primary.}
\label{LC}
\end{figure*}

V380\,Cyg (HD\,187879, KIC\,5385723) is a bright ($V=5.68$)
double-lined detached eclipsing binary with two early-B type
components in a 12.426\,d eccentric ($e=0.23$) orbit. It is an
important test for massive star models, with a primary star about
to start the hydrogen shell burning phase and a secondary still in
the early part of the main sequence. V380\,Cyg has been the
subject of numerous studies, of which \citet{popgui1998},
\citet{guinan2000}, and \citet{pavlovski2009} are the most
extensive, to which we refer for details on the properties of the
system.

The interpretation of the stellar parameters of V380\,Cyg deduced
from the binary data is problematic in the sense that an extreme
core overshooting parameter value above 0.5 (in local pressure scale
heights) is needed to match them to theoretical evolutionary tracks
\citep{pavlovski2009}. Such a high core overshooting is not
supported by recent seismic estimates, which rather point to values
below 0.2 for single B-type pulsators with similar masses
\citep[e.g.,][]{aerts2003,aerts2011,pamyatnykh2004,briquet2011}.
Nevertheless, \citet{briquet2007} deduced an overshooting parameter
of $0.44\pm0.07$ for the $8.2\pm0.3$\,M$_\odot$ primary of the
detached spectroscopic binary $\theta\,$Oph.

Clearly, an independent seismic evaluation of the core overshooting parameter of
the components of V380\,Cyg is of major importance. However, despite the
position of the two binary components in the instability strips for B stars,
searches for oscillations in the binary so far led to null detections. We report
the discovery of intrinsic variability from uninterrupted high-precision space
photometry obtained by the {\it Kepler} satellite and interpret it in terms of
gravity ($g$-mode) oscillations of the primary.


\section{Observations}

\subsection{{\it Kepler} customized aperture photometry}

The case study of the prototypical pulsator RR\,Lyr, with a maximum brightness
of $V=7.06$ during its pulsation cycle, showed that {\it Kepler}'s instrument is
capable of performing high-precision photometry on saturated targets
\citep{bryson2010,kolenberg2011} by using a customized aperture mask.  A
dedicated mask was thus defined to collect all flux from V380\,Cyg and the
binary was successfully observed in short cadence (58\,s) mode during Quarter 7
(Q7) of the mission, for about 90\,d continuously. This led to the conclusion
that low-amplitude variability occurs, aside variability from the eclipses and
brightenings, so the star was reobserved during Quarter 9 (Q9).

The Q9 {\it Kepler} light curve is shown in Fig.\,\ref{LC}.
Variability on timescales shorter than the orbital period is
obvious throughout the entire dataset, including during the
eclipses. The {\it Kepler} light curve is fully compatible with
but more precise than ground-based photometric data, which had
previously revealed scatter at the mmag level \citep{guinan2000}.

\subsection{HERMES high-resolution spectroscopy}

\tabcolsep=1pt
\begin{table}
  \caption{\label{tab:wd} Summary of the parameters
    for the {\sc wd2004} solution of the {\it Kepler} short-cadence light curve
    of V380\,Cyg. For further details on the control parameters we refer to
the
    {\sc wd2004} user guide \citep{wilvan2004}.
    A and B refer to the primary and secondary stars, respectively. }
\begin{tabular}{llc} \hline
Parameter             & {\sc wd2004} name & {Value($\pm$Error)}            \\
\hline
{\it Control parameters:} \\
{\sc wd2004} operation mode       & {\sc mode}            & {0}                \\
Treatment of reflection           & {\sc mref}            & {2 (detailed)}     \\
Number of reflections             & {\sc nref}            & {2}                \\
Limb darkening law                & {\sc ld}              & {1 (linear)}       \\
Numerical grid size (normal)      & {\sc n1, n2}          & {60}               \\
Numerical grid size (coarse)      & {\sc n1l, n2l}        & {40}    \\[3pt]
{\it Fixed parameters:} \\
Orbital period ($P_{\rm orb}$, d)                & {\sc period}          & {12.425719}        \\
Reference time of minimum (BJD)   & {\sc hjd0}            & {2441256.544}      \\
Mass ratio ($q$)                  & {\sc rm}              & {0.62}             \\
Effective temperature A ($T_{\rm eff}^{\rm A}$, K)       & {\sc tavh, tavc}      & {21500}            \\
Effective temperature B ($T_{\rm eff}^{\rm B}$, K)       & {\sc tavh, tavc}      & {22000}            \\
Rotation rates                    & {\sc f1, f2}          & {1.0}              \\
Gravity darkening                 & {\sc gr1, gr2}        & {1.0}              \\
Bolometric albedos                & {\sc alb1, alb2}      & {1.0}              \\
Bolometric LD coeff.\ A           & {\sc xbol1}           & {0.648}            \\
Bolometric LD coeff.\ B           & {\sc xbol2}           & {0.685}            \\
Third light                       & {\sc el3}             & {0.0}              \\
Passband LD coeff.\ B             & {\sc x2}    &   {0.262}\\[3pt]
{\it Fitted parameters:} \\
Phase shift                        & {\sc pshift} & -0.0493$\pm$0.0003\\
Star\,A potential                  & {\sc phsv}   & 4.62$\pm$0.05  \\
Star\,B potential                  & {\sc phsv}   &  11.1$\pm$0.6       \\
Orbital inclination ($i$, \degr)   & {\sc xincl}  &  80.5$\pm$0.5       \\
Orbital eccentricity ($e$)         & {\sc e}      &  0.235$\pm$0.005     \\
Longitude of periastron ($\omega$, \degr)    & {\sc perr0}  &  135$\pm$2         \\
Light from star\,A                 & {\sc hlum}   &  12.02$\pm$0.10      \\
Light from star\,B                 & {\sc clum}   &  0.777$\pm$0.011     \\
Passband LD coeff.\ A              & {\sc x2}     & 0.08$\pm$0.03\\[3pt]
{\it Derived parameters:} \\
Fractional radius of star\,A ($R_{\rm A}/a$)       &           &    0.2655$\pm$0.0020    \\
Fractional radius of star\,B ($R_{\rm B}/a$)       &           &
0.0648$\pm$0.0011
\\[3pt]
{\it Physical properties:} \\
Mass of star\,A ($M_{\rm A}$, $M_\odot$)         &            &     11.80$\pm$0.13      \\
Mass of star\,B ($M_{\rm A}$, $M_\odot$)         &           &     7.194$\pm$0.055     \\
Radius of star\,A ($R_{\rm A}$, $R_\odot$)       &           &     16.00$\pm$0.13      \\
Radius of star\,B ($R_{\rm B}$, $R_\odot$)       &          &     3.904$\pm$0.067     \\
Surface gravity star\,A (log$g$)    &            &     3.102$\pm$0.007     \\
Surface gravity star\,B (log$g$)    &            &     4.112$\pm$0.015     \\
Orbital semimajor axis ($a$, $R_\odot$)  &           &     60.25$\pm$0.19      \\
$\log(L_{\rm A} / L_\odot)$      &            &    4.711$\pm$0.023     \\
$\log(L_{\rm B} / L_\odot)$      &            &     3.474$\pm$0.047     \\
Distance ($d$, pc)                       &            &       987$\pm$17        \\
\hline
\end{tabular}
\end{table}

Given the discovery of low-amplitude variability in the {\it Kepler\/} Q7 data,
we initiated a ground-based spectroscopic campaign. We acquired an extensive
time-series of high-resolution, high signal-to-noise (S/N) spectra, starting
during {\it Kepler\/} Q9 and lasting for 142 days. The spectra were taken with
HERMES, the fibre-fed high-resolution spectrograph on the Mercator telescope
(Observatorio del Roque de los Muchachos, La Palma, Canary Islands), which
samples the entire optical range (380--900\,nm) with a resolution of 85\,000
\citep{raskin2011}. We obtained 150 observations during out-of-eclipse phases
and 256 exposures during primary or secondary eclipses. The typical exposure
time was 1200\,s. The data taken simultaneously with the {\it Kepler\/} Q9
observations is represented on Fig.\,\ref{LC}.

\begin{figure}
\begin{center}
\rotatebox{270}{\resizebox{5.5cm}{!}{\includegraphics{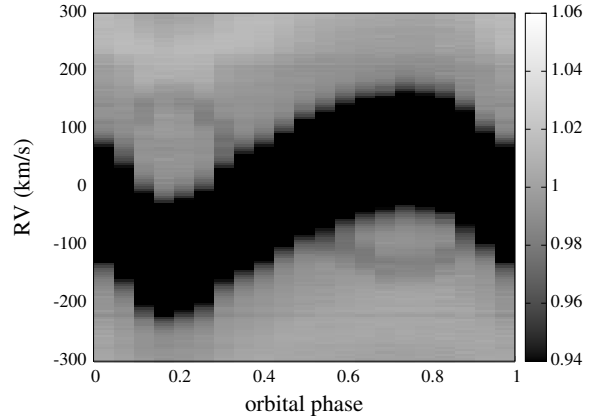}}}
\end{center}
\caption{Gray-scale representation of the LSD profiles folded
according to the orbital period of 12.425719\,d. Phase zero
corresponds to the time of primary minimum as derived by
\citet{guinan2000}.}
\label{gray}
\end{figure}

The reduction of the spectra has been carried out using the dedicated HERMES
software pipeline, including bias subtraction, cosmic ray filtering, flat
fielding, wavelength calibration using a ThAr lamp, and order
merging. Normalization to the local continuum was done manually by fitting a
spline function to carefully selected continuum points. We computed Least
Squares Deconvolution (LSD) profiles following \citet{donati1997}. These are
shown in a gray-scale representation according to the orbital period of $P_{\rm
  orb}=12.425719$\,d \citep{guinan2000} in Fig.\,\ref{gray}. The double-lined
nature of the binary is clearly recovered in our high-quality spectroscopy.

\section{Improved analysis of the binarity}

\subsection{Spectral disentangling}

We applied the technique of spectral disentangling ({\sc spd}) in velocity
(Fourier) space as implemented in the {\sc FDBbinary} code \citep{ilijic2004},
using three spectral intervals centred on strong helium or metal lines
(He\,{\small I} 4471\,\AA\ and Mg\,{\small II} 4481\,\AA; He\,{\small I}
4920\,\AA; the Si\,{\small III} triplet at 4550--4575\,\AA).  The orbital
elements we deduced are the time of periastron, $T_{\rm peri}$, the
eccentricity, $e$, the longitude of periastron, $\omega$, and the amplitudes of
the RV variations for the primary (star\,A) and secondary (star\,B) components,
$K_{\rm A}$ and $K_{\rm B}$.

Since the effective temperatures ($T_{\rm eff}$) of the two stars are similar
\citep{guinan2000,pavlovski2009}, their spectra are also similar at optical
wavelengths.
The orbital solutions for all three considered
short spectral intervals led to a consistent set of values: $T_{\rm
  peri}=54602.824\pm0.161$\,d, $e=0.2338\pm0.0026$,
$\omega=139.54\pm0.58$\,degrees, $K_{\rm A}= 94.266\pm0.034$\,km\,s$^{-1}$ and
$K_{\rm B}=154.59\pm0.31$\,km\,s$^{-1}$. The quoted errors are the standard
deviations of the {\sc spd} solutions for the three selected wavelength ranges.

\subsection{Light curve modelling}

The orbital effects in the light curves were modelled using the
Wilson-Devinney code \citep{wildev1971} in its 2004 version
(hereafter {\sc wd2004}), with automatic iteration performed using
the {\sc jktwd} wrapper \citep{southworth2011}.  We converted the
data to orbital phase using the ephemeris of \citet{guinan2000},
sorted them and reduced them into 202 binned datapoints. The bins
cover the eclipses five times finer than other phases, but are
otherwise equally spaced.

A variety of possible solutions were tried, with different numbers
of fitted parameters and model options. The full set of input,
output and control quantities for our final solution is given in
Table\,\ref{tab:wd} and the fit is shown as a full line in the
upper panel of Fig.\,\ref{LC}. The fit was performed in Mode 0,
which renders the effective temperatures inconsequential by
decoupling them from the solution. The detailed treatment of
reflection was found to be necessary for V380\,Cyg. We
assumed gravity darkening coefficients and albedos of unity. Third
light was found to be negligible so was set to zero. A linear limb
darkening law was used, and bolometric coefficients for both stars
plus the $K_{\rm p}$-band coefficient for star\,B were fixed to
values obtained using the tables of \citet{vanhamme1993}.
The uncertainties on the photometric parameters listed in Table\,\ref{tab:wd}
come from the scatter of appropriate solutions with different sets of fitted
parameters, and are much greater than the formal errors of the best single
solution.

\begin{figure*}\begin{center}
\includegraphics[scale=0.70]{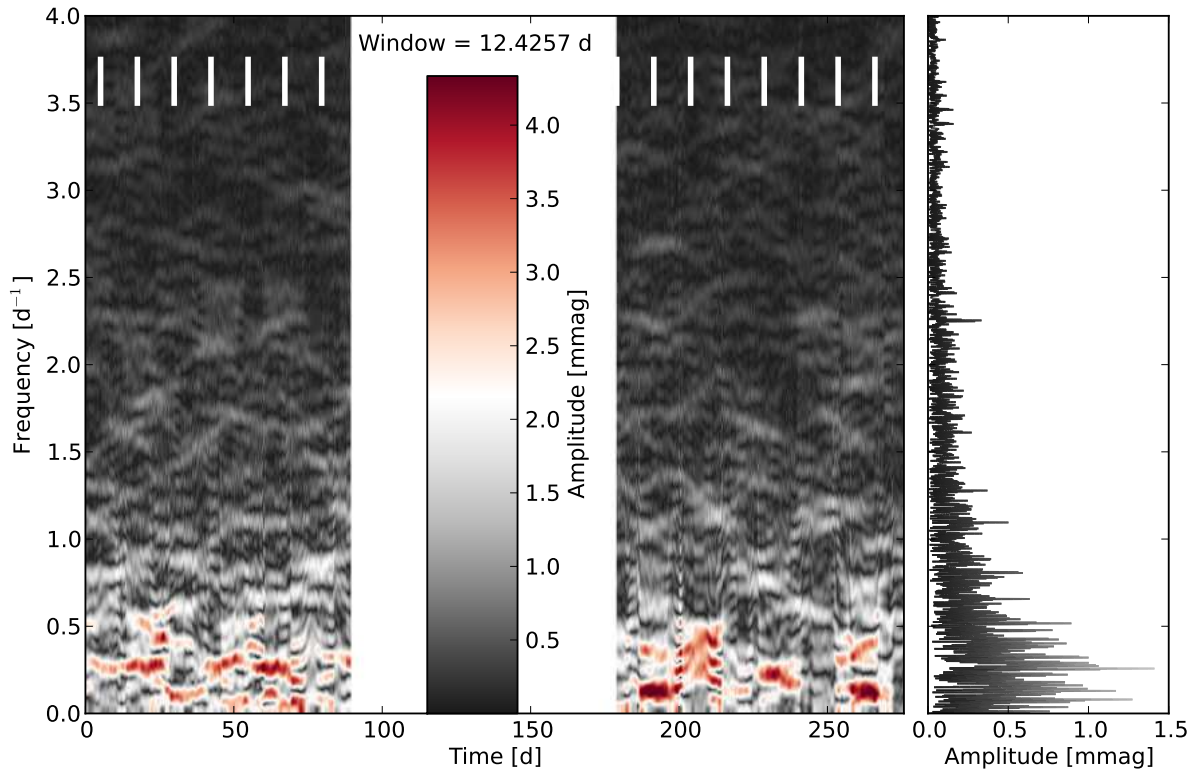}
\end{center}
\caption{Left: Short Time Fourier Transform of the residual light
curve shown in
  the middle panel of Fig.\,\protect\ref{LC}, for a window width of $P_{\rm
    orb}$. Right: summed amplitude spectrum. Thick white lines near the top of panels indicate the position of periastron passages.}
\label{STFT}
\end{figure*}

For our adopted solution we fitted for a phase shift with respect
to the orbital ephemeris, the potentials and $K_{\rm p}$-band
light contributions of the two stars, the $K_{\rm p}$-band limb
darkening coefficient of the primary star (star\,A), $e$, $\omega$
and the orbital inclination. These parameters are given in the
lower part of Table\,\ref{tab:wd} alongside uncertainty values
obtained by comparing all preliminary solutions providing a good
fit with physically reasonable parameter values. We also give the
fractional radii (stellar radii divided by the orbital semimajor
axis) which are needed to calculate the physical properties of the
two stars. There is a good agreement with the parameters found by
\citet{pavlovski2009}.
Table\,\ref{tab:wd} also contains the full physical properties of the V380\,Cyg
system, calculated using the {\sc absdim} code \citep{southworth2005}, our new
spectroscopic and photometric results, and the $T_{\rm eff}$ values found by
\citet{pavlovski2009}. For distance we quote the value obtained using the 2MASS
$K_s$ apparent magnitude, bolometric corrections from \citet{bessell1998}, and
adopting a reddening of $E(B-V)=0.21\pm0.03$\,mag.

We estimated the projected rotational velocity of V380\,CygA from a
spectrum taken at periastron and one at apastron, taking into
account rotational and thermal broadening but ignoring pulsational
broadening --- see Sect.\,\ref{pulsations}.  The values are
91.5$\pm$7.0\,km\,s$^{-1}$ and 96.5$\pm$7.0\,km\,s$^{-1}$,
respectively, and point to a higher rotation rate at apastron than
at periastron, although the two values are within their error bars.
In any case, none of the two values is compatible with synchronous
rotation, which would require 66.1$\pm$1.1\,km\,s$^{-1}$ for the
radius we found. Assuming the rotational axis to be perpendicular to
the orbital plane, not necessarily the case in an eccentric
non-synchronised binary, we estimate $f_{\rm rot}$ to be
0.115\,d$^{-1}$ at periastron and 0.121\,d$^{-1}$ at apastron.


\section{The intrinsic variability of the primary}
\label{pulsations}
\tabcolsep=4pt
\begin{table}
\label{freqs}
\caption{Results of the frequency analysis from the various diagnostics after
  orbital subtraction. The Rayleigh limit amounts to 0.0036\,d$^{-1}$ for the
  photometry and 0.0071\,d$^{-1}$ for the spectroscopy. }
\begin{center}
\begin{tabular}{ccccc}\hline
Diagnostic  & Frequency & Factor & Amplitude & Significance \\
&  (d$^{-1}$) & $f_{\rm orb}$ & & level (S/N) \\
\hline
{\it Kepler\/} LC & 0.2572 & 3.20 & 1.40 & 8.0\\
(mmag) & 0.0802 & 1.00 & 1.25 & 7.1\\
& 0.1295 & 1.61 & 1.15 & 6.6\\
& 0.2769 & 3.44 & 1.02 & 5.8\\
& 0.3204 & 3.98 & 1.00 & 5.7\\
& 0.1611 & 2.00 & 0.98 & 5.6\\
& \vdots & \vdots & \vdots & \vdots \\
& 3.4611 & 43.01 & 0.16 & 3.6\\
& 3.3802 & 42.00 & 0.16 & 3.5\\
& 2.2535 & 28.00 & 0.31 & 3.5\\
& 0.8053 & 10.01 & 0.56 & 3.2\\
\hline
$<v>$ & 0.0800 & 0.99 & 1.64 & 8.5 \\
(km\,s$^{-1}$) & 0.8062 & 10.02 & 1.17 & 6.1 \\
& 0.8853 & 11.00 & 0.94 & 4.9 \\
& 0.2738 & 3.40  & 0.80 & 4.2 \\
& 1.3682 & 17.00 & 0.57 & 3.0 \\
\hline
Pixel-by-pixel & 0.8868 & 11.02 & 0.0104 & 5.2\\
(continuum units) & 0.0350 & 0.43 & 0.0118 & 3.8\\
& 0.0797 & 0.99 & 0.0099 & 3.2\\
\hline
\end{tabular}
\end{center}
\end{table}

\begin{figure}\begin{center}
\rotatebox{0}{\resizebox{8cm}{!}{\includegraphics{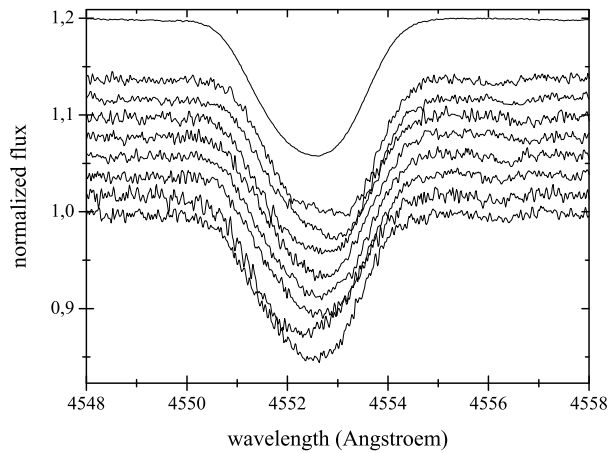}}}
\end{center}
\caption{Snapshots of the Si\,{\small III} line and its disentangled version for
  star\,A shifted to its center of gravity after subtraction of the disentangled
  profile of the secondary.}
\label{LPVs}
\end{figure}

The residuals of the light curve of both Q7 and Q9 (see middle panel of
Fig.\,\ref{LC} for Q9) were subjected to frequency analysis after detrending.
The outcome in the region $[0,4]\,$d$^{-1}$ ($[0,46]\,\mu$Hz) is shown in the
right panel of Fig.\,\ref{STFT}. Strong excess power is present at numerous
frequencies below 1\,d$^{-1}$ (11.57\,$\mu$Hz), the highest amplitude signals
being very significant --- see Table\,\ref{freqs}, where the significance level
was computed from the local average amplitude in the spectrum over a range of
3\,d$^{-1}$.  In addition, isolated peaks occur in $[1,4]\,$d$^{-1}$.  Most of
the significant frequencies are multiples of the orbital frequency, while others
are not.
The Short Term Fourier Transform computed for a window width taken
equal to the orbital period (left panel of Fig.\,\ref{STFT}) shows that part of
the variability damps and re-appears throughout the data set.

We also searched for residual variability in the strong Si\,{\small
III} line of the primary after {\sc spd} (Fig.\,\ref{LPVs}).  We
computed the line moments following the definition of
\citet{aerts2010} of all 406 spectra, corrected for the disentangled
profile of the secondary and shifted to the center of gravity of the
primary.  The $\langle{v}\rangle$-values obtained in this way for
the spectra taken simultaneously with the {\it Kepler\/} Q9
photometry are shown in the lower panel of Fig.\,\ref{LC} while the
frequencies found in this independent data set are also listed in
Table\,\ref{freqs}.  A pixel-by-pixel analysis \citep{zima2008} was
also done. Comparing all frequency analyses results in
Table\,\ref{freqs} leads to the conclusion that the line-profile
variability of the primary is compatible with the {\it Kepler\/}
residual photometry in the sense that power is found at multiples of
$f_{\rm orb}$ and at some other frequencies.  The frequencies in the
disentangled Si\,{\small III} lines cannot be due to the
prewhitening of the orbital motion, because the variability occurs
as {\it asymmetries\/} in the profiles and are thus intrinsic to the
primary (Fig.\,\ref{LPVs}).


\section{Discussion and Interpretation}

The intrinsic photometric and line-profile variability detected in
the residual data of V380\,Cyg\,A after orbital subtraction are
naturally explained in terms of gravity-mode oscillations. Other
causes of the intrinsic variability, such as migrating spots or
rotational modulation, are unlikely as they would require a complex
time-varying surface pattern and one would expect a much more
prominent presence of the rotational frequency in that case. We also
found no evidence of mass loss in the spectroscopy.

Some of the intrinsic frequencies detected in V380\,Cyg\,A are
connected with the orbital frequency. Given their time-dependent
amplitudes (Fig.\,\ref{STFT}) we suspect that they are tidally
triggered at periastron while damped further in the orbit. If
confirmed, this finding holds the potential to model the interior
structure of the primary in great detail and to deduce a seismic
estimate of the core overshooting parameter which can then be
compared with the one deduced from isochrone fitting.  However,
additional data are needed to deduce if the amplitude and frequency
behaviour is indeed repetitive for numerous orbits. Such a situation
was found for few close binaries so far. One case is HD\,174884, an
eccentric short-period binary consisting of two cool B stars with
residual variability at low-order multiples of the orbital frequency
and two tidal oscillations induced at 8 and 13\,$f_{\rm orb}$
\citep{maceroni2009}. Resonant mode locking at 90 and 91\,$f_{\rm
orb}$ occurs in the highly eccentric 42\,d binary HD\,187091
consisting of two A stars \citep[KOI-54,][]{welsh2011}. In contrast
to these two cases, V380\,CygA exhibits an excess bump of power
rather than only well-isolated stable frequencies.  Future {\it
Kepler\/} data will cover three more quarters while additional
spectroscopy will be gathered to unravel the pulsational behaviour
of this massive binary.

\section*{acknowledgements}
Funding for the {\it Kepler} Discovery mission is provided by NASA's Science
Mission Directorate.  The authors gratefully acknowledge the {\it Kepler} and
Mercator teams, whose outstanding efforts have made these results possible, and
Dr.\ Yves Fr\'emat for carrying out part of the observations with the Mercator
telescope. The research leading to these results has received funding from the
European Research Council under the European Community's Seventh Framework
Programme (FP7/2007--2013)/ERC grant agreement n$^\circ$227224 (PROSPERITY).  JS
acknowledges financial support from STFC in the form of an Advanced
Fellowship. PD and BD are Postdoctoral and Aspirant Fellow of the FWO,
respectively. JD is funded by the Belgian federal science policy office
Belspo. PN is Boursier F.R.I.A., Belgium.

\end{document}